\documentclass[aps,pre,floats,twocolumn,showpacs,superscriptaddress]{revtex4-1}
\usepackage{amsmath}
\usepackage{graphicx}
\usepackage{epstopdf}
\usepackage{amsfonts}
\usepackage[utf8]{inputenc}
\usepackage{color}
\usepackage{cancel}
\usepackage{hyperref}% add hypertext capabilities
\usepackage[english]{babel}
\usepackage[dvipsnames]{xcolor}

\begin{document}

\title{Influence of topology in the mobility enhancement of pulse-coupled oscillator synchronization.}

\author{A. Beardo}
\author{L. Prignano}
\affiliation{Complexity Lab Barcelona. Departament de F\'{\i}sica de la Mat\`eria Condensada, Universitat de Barcelona, 08028 Barcelona, Spain}
\affiliation{Universitat de Barcelona Institute of Complex Systems (UBICS), Universitat de Barcelona, 08028 Barcelona, Spain}
\author{O. Sagarra}
\affiliation{DRIBIA Data Research, Carrer Roc Boronat 117, 08018 Barcelona, Spain}
\affiliation{Complexity Lab Barcelona. Departament de F\'{\i}sica de la Mat\`eria Condensada, Universitat de Barcelona, 08028 Barcelona, Spain}
\author{A.  D\'iaz-Guilera}
\affiliation{Complexity Lab Barcelona. Departament de F\'{\i}sica de la Mat\`eria Condensada, Universitat de Barcelona, 08028 Barcelona, Spain}
\affiliation{Universitat de Barcelona Institute of Complex Systems (UBICS), Universitat de Barcelona, 08028 Barcelona, Spain}

\begin{abstract}
In this work we revisit the nonmonotonic behavior (NMB) of synchronization time with velocity reported for systems of mobile pulse-coupled oscillators (PCOs). 
We devise a control parameter that allows us to predict in which range of velocities NMB may occur, also uncovering the conditions allowing us to establish the emergence of NMB based on specific features of the connectivity rule. 
Specifically, our results show that if the connectivity rule is such that the interaction patterns are sparse and, more importantly, include a large fraction of non reciprocal interactions, then the system will display NMB.  
We furthermore provide a microscopic explanation relating the presence of such features of the connectivity patterns to the existence of local clusters unable to synchronize, termed frustrated clusters, for which we also give a precise definition in terms of simple graph concepts. We conclude that, if the probability of finding a frustrated cluster in a system of moving PCOs is high enough, NMB occurs in a predictable range of velocities.
\end{abstract}
\maketitle

\section{Introduction}
\label{sec_intro}

Complex systems are characterized by emergent properties that cannot be immediately inferred from the properties of the units forming it. Among these properties, synchronization has become one of the most paradigmatic examples, because synchronization processes are ubiquitous in nature and play a very
important role in many different contexts such as biology, ecology, climatology, sociology, and technology \cite{pikovsky2003synchronization,osipov2007synchronization,boccaletti2008synchronization}. Periodic interactions between the system units lead to a common rate of entrainment, which can be characterized either by a common phase or by a common frequency. Similarity between the periods of the units is crucial to achieve such a synchronized state but there is another ingredient that plays also a very significant role: the pattern of interactions between the units \cite{arenas2006synchronization,arenas2008synchronization,dorfler2014survey}. It is important not only for determining the time scale to reach a stationary state but in some cases -as in the situation under study in the present work- it can even prevent the synchronization of identical units \cite{PSD13,perez2017control}.

%intro general: IFOS
In the past few years, renewed interest has emerged in the study of systems of coupled oscillators that move in space, forming complex time-dependent networks. Such setups can be used as simplified representations of real (more complex) systems to study the efficiency and feasibility of communication protocols among its units. 

These models, despite their apparent simplicity, display a variety of properties that cannot be explained based on an aggregation of the characteristics of the elements forming the system, but emerge from the interaction patterns themselves and their rules of change.

Prominent examples where this modeling can be useful range from technological applications (groups of autonomous self-propelled vehicles) \cite{baronchelli2012consensus,BFFF16} to the study of synchronization in ethology (anurans, bush crickets and fireflies) \cite{chicoli2016fish,swain2015fish,giardina2008,deisboeck2009}. 
Also, mapping the mobility of the units to a certain change in their environment (who or what you see/follow/interact with at a given moment) can be used to study social phenomena and even unexpected financial behaviors \cite{saavedra2011}

% moving IFOS
While the emergence of such behaviors is by no means restricted to systems with moving units, recently, interesting and intriguing phenomena triggered by the motion of its constituents have been studied. 
In particular, studies have been performed to describe how the ability of a system of coupled oscillators to achieve a synchronized state is affected by the speed of their motion under different experimental conditions and settings.
Generally speaking, moving faster usually makes the time the system needs to reach a coherent state shorter \cite{fujiwara2011km,fujiwara2016chaos,levis2017prx}
%\tp{Initial citations here! - Naoya, Kurtz, PRL, etc...}

% Surprising PRL
Nevertheless, more recent studies have shown how this is not always the case. When the coupling is highly nonlinear [i.e., for pulse-coupled oscillators (PCOs), also called integrate-and-fire oscillators (IFOs)] it may happen that increasing the velocity is not beneficial for the achievement of a synchronized state.
It has been suggested~\cite{PSD13} that two ingredients are necessary for this behavior to be displayed: the interaction pattern has to be (a) sparse and (b) nonreciprocal.

% Conditions for sync
The first condition means that each oscillator is limited to interact with just very few units at the same time. Thus, without motion the system is disconnected -below the static percolation threshold~\cite{dall2002rgg} - and hence unable to synchronize globally because no signal can propagate through the entire system. 
Therefore, mobility is necessary to achieve synchronization.
The second condition refers directly to the details of the interaction rule. It must include a certain degree of asymmetry.
In summary, if moving pulse-coupled oscillators are (a) allowed to receive and send a signal only to a few other nearby elements that (b) may or may not correspond to them depending on a nonsymmetric interaction rule, then the synchronization time has a nonmonotonous dependency  on the velocity of motion.
That is, a velocity increase does not always correspond to a decrease in the system's synchronization time.

For these setups, broadly three possible scenarios have been identified:
(1) for slow speeds, moving a little faster promotes synchrony in a shorter time;
(2) for fast enough velocities, the synchronization time approaches a minimum constant value which becomes independent of the speed of motion;
and (3) for intermediate values, when the velocity is increased, counterintuitively, the system takes longer on average to reach a coherent state, sometimes being completely unable to synchronize.
% -thus taking “infinite time” in computational terms-.

The underlying hypothesis to explain this phenomenology is that different synchronization mechanisms are at work for the two extreme regimes. In case 1, synchrony is achieved at the level of small groups of units that are able to transmit information among themselves. These small groups synchronize internally and then break and recombine into new groups with increasingly less diverse phases. This iterative process leads the system to synchronize at a global scale through a sort of coalescent process. In case 2, every oscillator has the chance to interact with many others in a short time span, so global synchrony emerges directly through individual interactions rather than by repeated cluster recombination.

This description is very general and holds for any kind of coupling. For instance, it has been proven correct for moving Kuramoto oscillators \cite{fujiwara2011km}.
However, if the coupling is highly non linear as in PCOs, difficulties arise for intermediate velocities when the typical time between two consecutive changes in the interaction pattern is comparable with the time that local groups take to synchronize. Then, some clusters are broken before they synchronize but at the same time the interactions are not rewired fast enough. In this case, the two typical timescales of the system, that of the motion of its units and that of the synchronization of local clusters, may interfere in a very ineffective way.

This hypothesis has been demonstrated for a particular minimal model with particulary simple interaction rules~\cite{PSD13} where a semianalytic estimation of the value of the limiting velocities that separate the three regimes has been proposed.

However, nonmonotonic behavior has been observed in other settings. In particular, more recent works~\cite{PRG15,perez2017control} have confirmed that a sparse and nonreciprocal interaction pattern is a necessary condition for such behavior to be observed, yet the validity of the general interpretation based on the two timescales has not been verified.

% What is done in the paper
In this paper, we analyze the model proposed in~\cite{PRG15,perez2017control} showing how it fits the interpretation proposed in~\cite{PSD13}. Additionally, we introduce a general explanation of what the unfruitful interplay between timescales is and how and why it is related to features (a) and (b) of the interaction pattern.
% in term of the topological characterization of local clusters.

In Sec. \ref{sec_model}, we roughly describe the model under study. We then determine for this specific model the velocity at which the expected time between two consecutive changes in the interaction pattern is the same as the average time local clusters need to synchronize. We show how, starting from just above this precise value of the speed, the dependency  of the synchronization time on the velocity of the oscillators may change depending on whether conditions (a) and (b) are satisfied. 
We thus validate the hypothesis about the ineffective interplay between the two time scales for an additional, more realistic (suitable to be implemented with real robots), experimental setting than the one studied in~\cite{PSD13}.

In the second part, Sec. \ref{sec_conditions}, we focus on studying the relationship between conditions (a) and (b) and the appearance of the non monotonic behavior through the analysis of the oscillators interaction patterns. In particular, we show how the existence of local configurations that are not able to reach a synchronized local state is what makes the interplay between the time scale fruitless.

To conclude, we present a complete explanation of the necessary and sufficient conditions for this peculiar and unexpected phenomenon to occur in terms of the microscopic topological and dynamical characterization of the system.

% fig 0: Model schema
\begin{figure}[htbp]
\begin{center}
\includegraphics[width=0.4\textwidth]{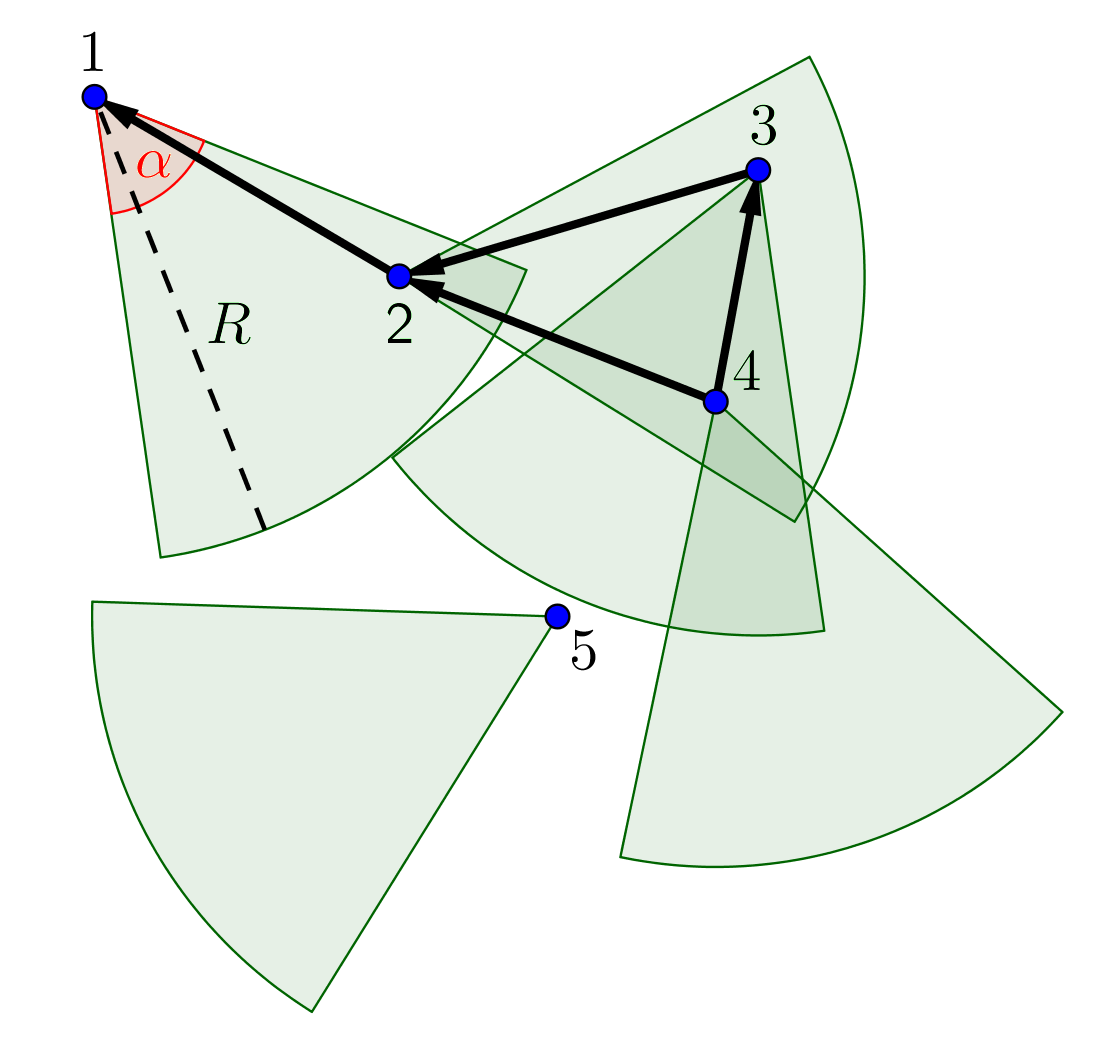}
\caption{Illustration of the interaction rule. The arrows indicate the neighbors of each oscillator and the cones of vision are the shaded green areas determined by $\alpha$ and $R$. Upon firing by an oscillator, its neighbors are affected by a phase update.}
\label{fig0}
\end{center}
\end{figure}

\section{The ineffective interplay of two time scales.}
\label{sec_model}

The model considered in this paper was introduced in~\cite{PRG15,perez2017control} and can be regarded as a modified version of the minimal model originally proposed in~\cite{PSD13}.

It consists of a population of $N$ moving oscillators with velocity $V$ and random orientation on a square of side length $L$ with finite boundary conditions. When a unit reaches a border, its motion is reoriented randomly inside the box.

The internal phases of the agents $\phi \in (0,1)$ increase uniformly with frequency $\tau^{-1}$,
\begin{align}
\frac{d\phi_i}{dt}=\frac{1}{\tau} \qquad \forall \,i = 1,\dots, N,
\end{align}
until they reach a maximum value of 1, when a firing event occurs and the phase is reset. 

The interaction rule of this particular model is based on cones of vision (COV), which are circular sectors centered in the oscillators and oriented in the direction of their motion (parallel to $V$). The COV are characterized by a radius $R$ and an angle $\alpha$ that are the same for all the units in the system (see Fig.~\ref{fig0}). 
Whenever a firing event is triggered, all units that have the emitting oscillator inside their COV are affected.
Upon a firing by oscillator $i$ at time $t$, all such oscillators $n$ - which we call neighbors of unit $i$ from now onward - receive a signal and update their phases $\{\phi^i_n\}$ by a factor $\varepsilon$:
\begin{align}
\phi_{i}(t^-)=1 \Rightarrow 
\left \{ 
\begin{array}{l} 
\phi_i(t^+)=0\\
\phi^i_n(t^+)=(1+\varepsilon)\phi^i_{n}(t^-) 
\end{array}
\right. .
\end{align}

The phase update is performed at frozen time until the phases of all oscillators have been updated (some agents may reach their threshold and fire upon receiving a phase update from a firing neighbor). Then the phases evolve again uniformly in time (we take $\tau=1$ to fix the time scale) until another firing is triggered. We consider that the system is synchronized when a succession of $N$ firing events takes place, or equivalently when all the oscillators have exactly the same internal phase.

Following~\cite{PSD13}, in order to verify the hypothesis that the nonmonotonic behavior (NMB) is caused by an ineffective interplay between the two time scales that characterize the system, we must determine both the average time between two sequential changes in the interaction pattern, $T_C$, and the average local synchronization time $T_L$. By local synchronization time we mean the time that a subset of interacting oscillators takes to reach equal phases. Such a subset is defined as follows: taking one unit as a starting point, the cluster includes all its neighbors (those that receive its signal) and the oscillators whose neighbor is this unit (that send their signal to it); then the same is done for every newly included unit until no new oscillator is added to the group. We call such subsets of oscillators local clusters and their definition corresponds to what in graph theory is called a weakly connected component of a direct graph. In Fig.~\ref{fig0}, oscillators $1$, $2$, $3$, and $4$ form a local cluster, while oscillator $5$ belongs to another one whose only element is oscillator $5$ itself.

If our hypothesis is correct, then a change in the dependency of the synchronization time on the velocity should be observed when these two characteristic times approach each other if conditions (a) and (b) of the interaction pattern are satisfied. In other words, there should exist a typical velocity - depending on the parameters of the system - such that $T_C$ is equal to $T_L$. Above this velocity, the system may exit the slow regime and the synchronization time will display a NMB.

Besides the usual parameters that characterize every model of this type - the number of oscillators in the system $N$, the coupling constant $\varepsilon$, and the size of the box $L$ - the model under study is defined by two additional parameters that determine the spatial details of the interaction rule: the reach (radius $R$) and shape (angle $\alpha$) of the COV. By varying these parameters it is possible to tune the average number of neighbors (through the area) and the proportion of nonreciprocal interactions (through the angle) thus directly affecting the properties of sparseness (a) and asymmetry (b) of the interaction pattern.

% fig 1: Tsync vs eta
% fig 2: Tsync vs eta
\begin{figure}[htbp]
\begin{center}
\includegraphics[width=0.5\textwidth]{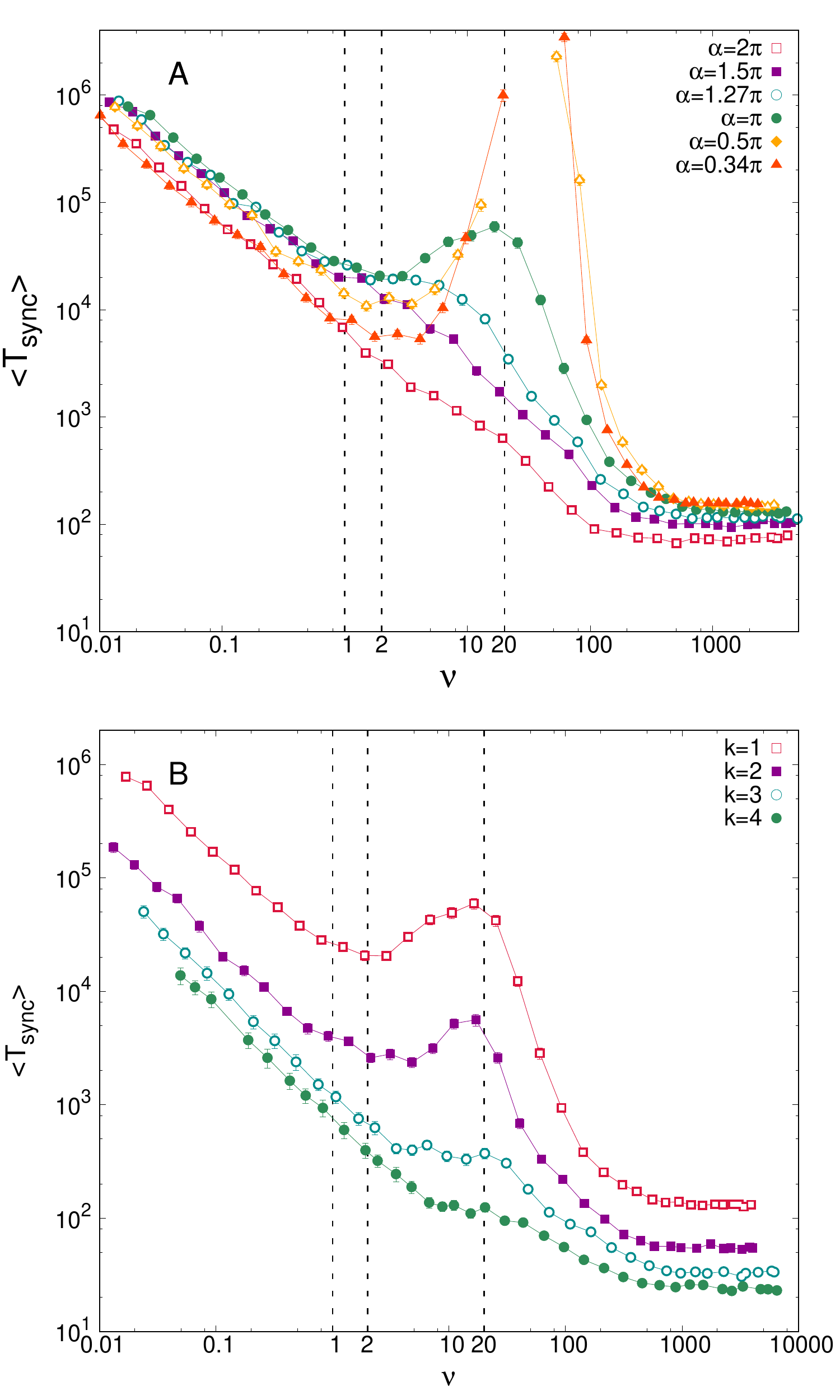}
\caption{$\langle T_{\text{sync}} \rangle$ vs $\nu$. (a) We fix $\bar{k}=1$. NMB is obtained by decreasing $\alpha$, or equivalently by enforcing the asymmetry of the interaction pattern. (b) We fix $\alpha=\pi$, so we consider partially asymmetric interactions. For this value of $\alpha$, the NMB is no longer observed if we increase the connectivity. 
%Therefore, we validate the necessary conditions (a) low connectivity and (b) asymmetric interactions for obtaining NMB in this particular model. 
In such cases where NMB is displayed,  $\nu=1$ (or equivalently $T_{L}=T_{C}$) characterizes the velocity from which the monotonic decreasing of $\langle T_{\text{sync}}\rangle$ is broken. 
%Hence we are able to confirm the hypothesis of ineffective interplay between similar time scales governing the local synchronization time $T_{L}$ and the changes in the interaction pattern $T_{C}$. 
$\langle T_{\text{sync}}\rangle$ is averaged over 75 realizations in a setting of $N=20$ oscillators in a box of side $L=200$, with coupling constant $\epsilon=0.1$. Hereafter, this same setting is implemented in all the figures.}
\label{fig12}
\end{center}
\end{figure}

In Fig.~\ref{fig12}(a) we show the average synchronization time $\langle T_{\text{sync}} \rangle$ as a function of the ratio $\nu= T_L/T_C$ for several values of $\alpha$ and a fixed area of the COV, fixing $N$, $L$ and $\varepsilon$ too. 
In order to keep everything but the fraction of nonreciprocal interactions fixed, the radius has been adjusted to force the average number of neighbors per oscillator to $\bar{k}=1$. 
The appropriated value of $R$ for each $\alpha$ can be computed analytically under periodic boundary conditions, yet, due to finite boundary effects, a correction needs to be applied on the simulations (see Appendix~\ref{app:scaling_k}).

The empirical value of $T_L$ for each set of parameters has been calculated numerically under static conditions, that is, for $V=0$. Initiating the system with random initial conditions (phases and orientation of the COV) several times, $T_L$ has been computed as the average time a reference oscillator takes to synchronize with the rest of the units in its local cluster. The cases in which it is not possible to achieve a coherent state after a fixed maximum number of cycles $T_{\text{max}} \gg \langle T_{\text{sync}} \rangle$ have been discarded assuming that they do not contribute to the general enhancement of coherency. Thus, $T_L$ is the average time synchronizable local clusters of oscillators of any size need to synchronize.

Concerning $T_C$ - the average time between two sequential changes in the connectivity pattern - notice that in principle it can be determined via semi-analytic calculations. Following the line of reasoning exposed in \cite{PSD13}, we can estimate $T_C$ as the time a unit needs to exit the COV of one of its neighbors, averaging over all the possible initial positions and orientations, divided by the total number of oscillators in the system. However, obtaining an explicit expression of this quantity as a function of $V$, $\alpha$ and $R$ is a tedious task that does not deserve the effort in this context. Hence, its value has been estimated numerically. In Appendix~\ref{app:pi_time} we show an example of how such calculation can be performed in the particular case of $\alpha=\pi$ to obtain the explicit dependency  $T_C(V,R)$.

Figure~\ref{fig12}(a) shows that, if $\alpha$ is small enough, starting from $\nu=1$ (that is, when $T_L = T_C$), the decreasing of $\langle T_{\text{sync}} \rangle$ slows down and then the appearance of NMB is patent for larger values of $\nu$ (see Appendix \ref{app_nu1} for further numerical evidences). The smaller the value of $\alpha$, the greater the deviation from the monotonous behavior. For larger values of the angle, the connectivity pattern gets increasingly symmetric and condition (b) is no longer satisfied, so the monotonic behavior is recovered. In particular, when $\alpha=2\pi$ we recover the fully symmetric model studied in~\cite{PSGD12} for which no deviation from the monotonous behavior has ever been observed, for any choice of the parameters.

%We have measured $T_C$ and $T_L$ via simulations for different values of $N$ and $\varepsilon$ and large enough $L$. 

%Contrary to the \tp{minimal} case~\cite{PSD13}, due to the great diversity of topological configurations obtained by varying $\alpha$ and $R$, approximating these times can only be done by computational means, however, for the particular case of $\alpha=\pi$ $T_C$ can be calculated semianalytically, see Appendix~\ref{app:pi_time}.  No es cierto. Se puede generalizar el cálculo analítico para calcular $T_C$ para cualquier $\alpha$. Nosotros fijamos $\alpha=\pi$ por simplicidad y para ver las dependencias en $V$ y $R$, ya que consideramos que es un caso representativo.}

In Fig.~\ref{fig12}(b) we plot again $\langle T_{\text{sync}} \rangle$ as a function of $\nu$ but this time for a fixed value of the COV angle, $\alpha=\pi$, with increasing radius of the cone in order to vary the average number of neighbors per oscillator. As expected, NMB can be observed for $\nu>1$ if $\bar{k}$ is small enough, while the nonmonotonicity fades for larger $\bar{k}$, in accordance with the results found in \cite{perez2017control}.
%, \tp{roughly above} the static percolation threshold~\cite{percolation}.
% (the point where a signal could flow among any pair of neighbors without the need for movement in the system). 

We are thus able to confirm that $\nu$ is indeed a suitable control parameter for this class of models, and not only for the particular example proposed in~\cite{PSD13}. 
For all the sets of parameters for which the system displays NMB, we observe a change in the dependency of the synchronization time above $\nu=1$. It then reaches a local minimum around $\nu=2$ and a peak roughly around $\nu=20$, which is when $T_L$ is comparable with $NT_C$ and almost no local cluster lasts long enough to be able to synchronize completely. These observations corroborate the hypothesis that it is actually the ineffective interplay of the two timescales that complicates the synchronization process. Moreover, it confirms that such phenomenology will be surely observed whenever two conditions are satisfied, that is, when both $\alpha$ and $\bar{k}$ are small enough.

Notice also that changing the coupling parameter $\varepsilon$ or the number of oscillators, $N$, does not affect the validity of our arguments (see Appendix~\ref{app:scaling_eps}).

%\textcolor{blue}{A relevant question in this case is how small must these parameters be. To quantify them, we use ... EXPLANATION OF ASYMETRY PARAMETERS AND SUCH.}
% fig 3: heatmap
\begin{figure}[htbp]
\begin{center}
\includegraphics[width=0.5\textwidth]{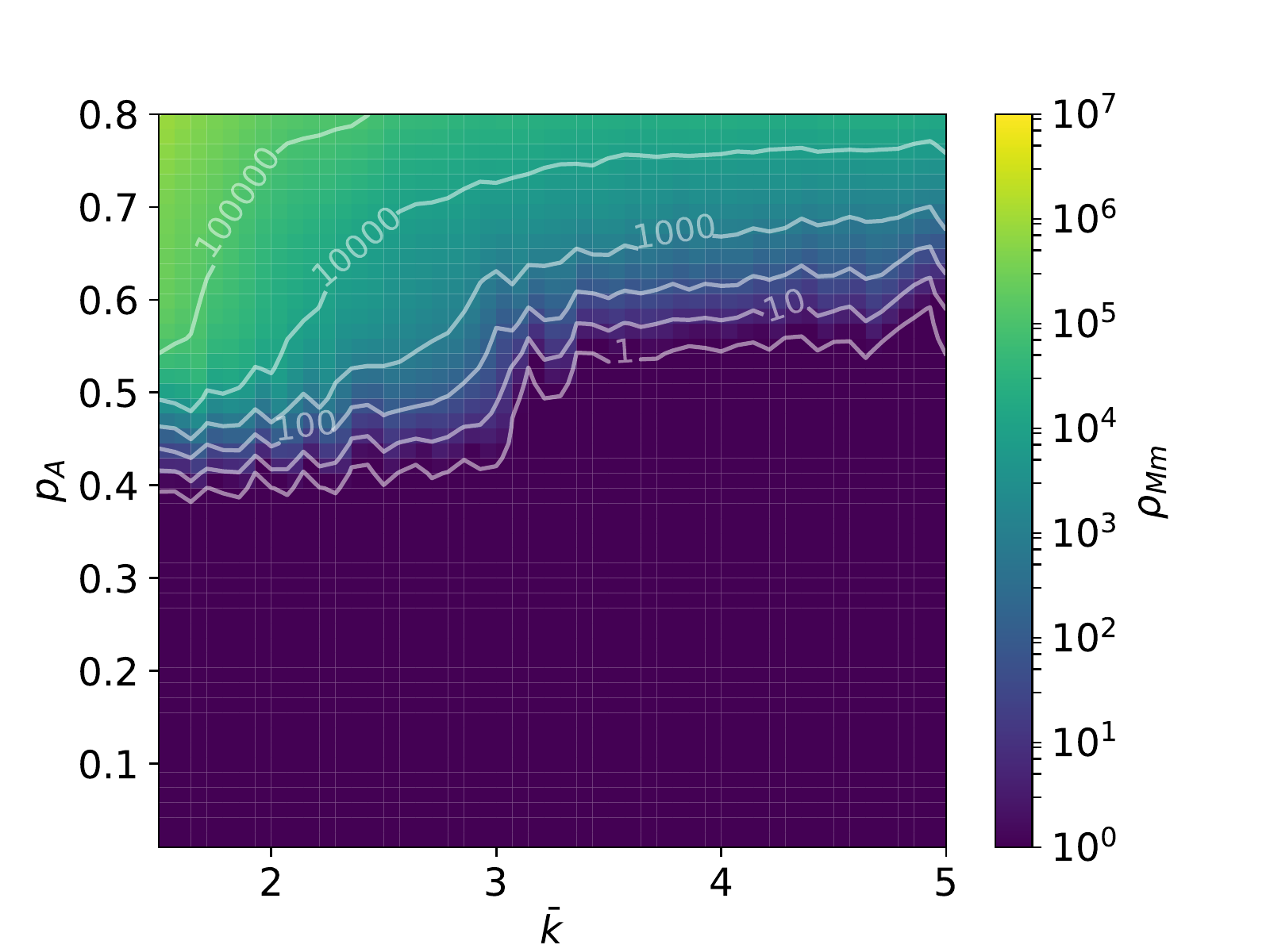}
\includegraphics[width=0.5\textwidth]{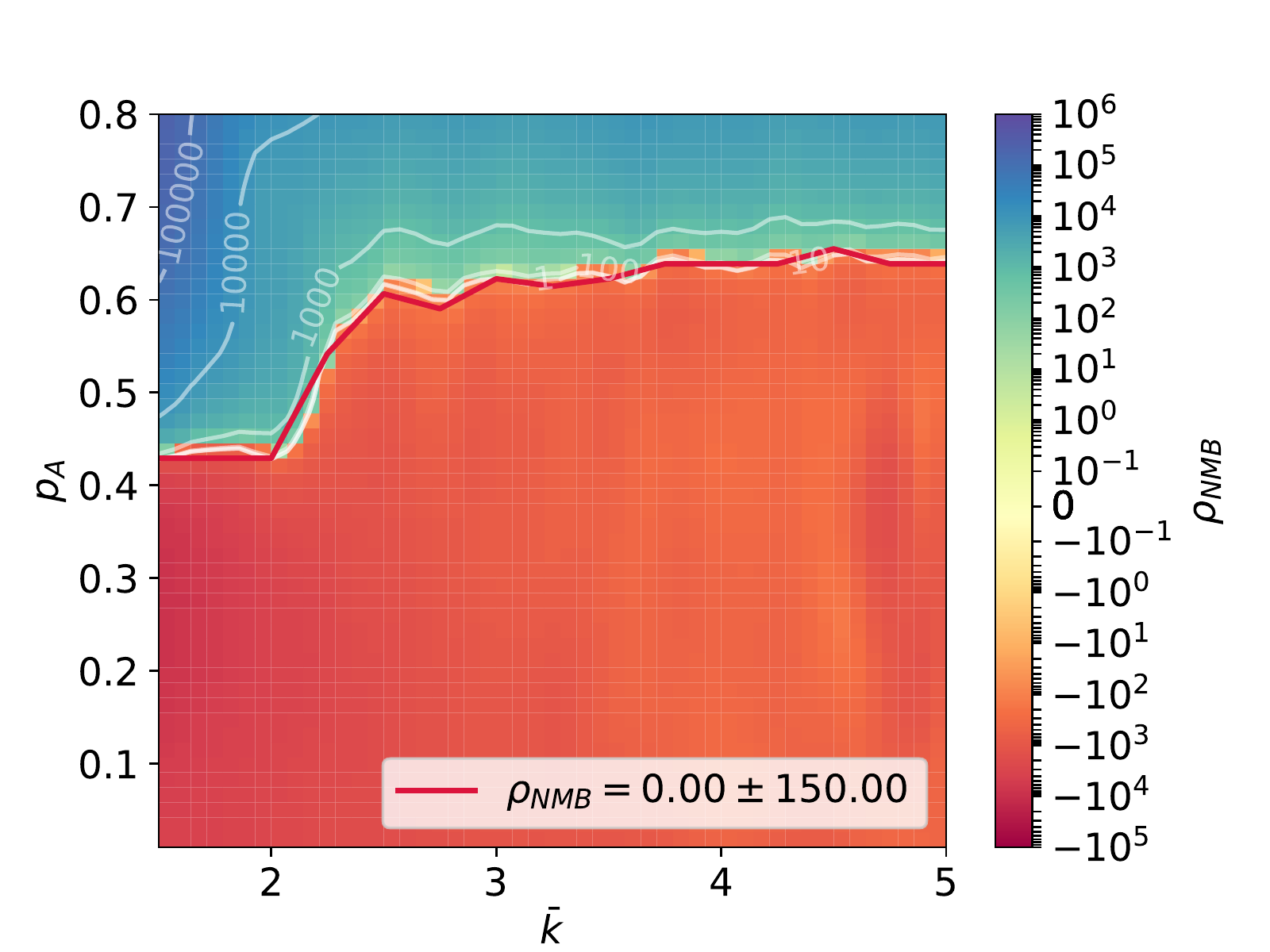}
\caption{ $\rho_{\text{Mm}}$ and $\rho_{\text{NMB}}$ vs $p_{A},\bar{k}$ with superimposed isolines. The boundary line represents the smoothed  interpolation of points for which $\rho_{\text{NMB}} \in [-150,150]$, which marks the appearance of NMB.  
Figures are averaged over 150 realizations with $N=20$ and $\epsilon=0.1$ and interpolated using a multiquadratic radial basis function (RBF). For visualization purposes, in the   panel, values for which $\rho_{\text{Mm}}<1$ have been saturated at $\rho_{\text{Mm}}=1$. In the lower panel, values are displayed using a \emph{symlog} scale, which is shown as $-\log |\rho_{\text{NMB}} |$ if $\rho_{\text{NMB}}<0$ with linear interpolation in the range $\rho_{\text{NMB}} \in [0.01,-0.01]$.}
\label{fig3}
\end{center}
\end{figure}

Let us introduce a metric able to capture in a quantitative way the degree of nonmonotonicity of the system behavior. 
For every pair ($\alpha$, $R$), we can compute the estimator $\rho_{\text{Mm}}=T_{\text{sync}}^{\text{MAX}}-T_{\text{sync}}^{\text{min}}$, that is, the the difference between the largest and the smallest value of $\langle T_{\text{sync}}\rangle$. 
More precisely, $T_{\text{sync}}^{\text{min}}$ is the first minimum of $\langle T_{\text{sync}}\rangle$ starting from $\nu=0$, hence, it may correspond either to a local minimum or to the asymptotic value for $\nu \to \infty$ (fast switching regime), depending on whether the system does or does not display NMB. 
Likewise, $T_{\text{sync}}^{\text{MAX}}$ is the maximum value of the average synchronization time for $\nu>\nu_{\text{min}}$, that is, for a value of $\nu$ larger than that corresponding to $T_{\text{sync}}^{\text{min}}$.

When the system behavior is monotonic, besides small fluctuations, there is no difference between $T_{\text{sync}}^{\text{MAX}}$ and $T_{\text{sync}}^{\text{min}}$, so $\rho_{\text{Mm}}\approx 0$. 
On the contrary, if the system displays a high degree of nonmonotonicity, the difference between the minimum and the maximum of $\langle T_{\text{sync}}\rangle$ is not negligible and $\rho_{\text{Mm}}$ takes increasingly larger values.

In the heatmap in Fig.~\ref{fig3}(a), $\rho_{\text{Mm}}$ is plotted against the average number of neighbors $\bar{k}$ and the expected fraction of nonreciprocal interactions $p_A=1-\alpha/2\pi$. Below $p_A\simeq 0.4$ the system behaves monotonically (dark blue). Above $p_A\simeq 0.4$, depending on the value of $\bar{k}$, it may display NMB. The transition between the monotonic and nonmonotonic regions is quite smooth, especially for relatively large $\bar{k}$.
According to the definition of $\rho_{Mm}$, when $\rho_{Mm}>1$, the system displays nonmonotonicity.
However, if $\rho_{Mm}<<T_{\text{sync}}^{\text{min}}$, the behavior can be classified as just slightly nonmonotonic: $\langle T_{\text{sync}}(\nu)\rangle$ displays a plateau for intermediate $\nu$, with a small bulge after $\nu=1$.
For larger $\rho_{Mm}$, we find stronger NMB, ranging from the ultra-NMB of the top left corner, where the sychronization time diverges for intermediate velocity, to the non diverging NMB displayed when the system is more densely connected but most of these connections are asymmetric (top-right part).
In the lower $\rho_{Mm}$ region the behavior of the system is perfectly uniform: increasing the velocity makes the synchronization time decrease monotonically.

Although this description captures all the relevant features of the system behavior, it does not provide a clear border between what is proper NMB and behaviors barely deviating from monotonicity, which do not suppose any relevant violation of the general rule that states that \textit{mobility enhances synchronization}.

%The existence of two clearly separated regions corresponding to the presence and absence of the NMB respectively, is even more evident when looking at Fig.~\ref{fig3}B. 
With the aim of differentiating properly nonmonotonicity from other trends, we propose a different estimator that takes positive values when the behavior is strongly nonmonotonic and negative ones otherwise.
Such an estimator is defined as the difference between the expected synchronization time near to the local minimum and to the local maximum: $\rho_{\text{NMB}}=\langle T_{\text{sync}}(\nu=20)\rangle - \langle T_{\text{sync}}(\nu=2)\rangle$. In this way, we are regarding as NMB only those cases such that increasing the velocity by a factor of 10 does not benefit the ability of the system to synchronize in a shorter time. 

The choice of the reference values of $\nu$ is, to some extent, arbitrary. Nonetheless, we have already shown that $\nu$ is a good control parameter for this class of systems and the average synchronization time as a function of $\nu$ in the NMB region displays common features, one of the most relevant being a common localized range of values at which the synchronization time slows down its decreasing, reaches a minimum, and subsequently starts to increase up to a maximum or a divergence. Hence, a slight change of the chosen values of $\nu$ does not modify the result shown in Fig.~\ref{fig3}(b): the existence of two clearly separated regions which correspond to the presence (blue) or absence (orange) of strong NMB, respectively.

In the figure, the border is highlighted by a red line. Above this line, the system displays a degree of nonmonotonicity that ranges from the ultra-NMB of the top left corner, where the synchronization time diverges for intermediate velocities, to the non diverging NMB displayed when the system is more densely connected but most of these connections are asymmetric (top right part). In all these circumstances, it is correct to affirm that there exists a range of velocities such that increasing the mobility of the oscillators makes the synchronization process considerably slower.

More specifically, we observe that the deviation from the monotonic behavior is maximal ($\rho_{\text{NMB}}>10^4$, dark blue) when $\bar{k}$ is very small, namely, $\bar{k}<1.8$, and $p_A$ is larger than $0.6$, that is, when conditions (a) and (b) are both satisfied. It is also worth noticing that when the fraction of asymmetric interactions is large enough ($p_A>0.6$), the strong NMB does not disappear completely by merely increasing the connectivity 
($10^2<\rho_{\text{NMB}}<10^4$).
In other words, it is not possible to affirm that a minimal connectivity is still a necessary condition for NMB when almost all the interactions are no-reciprocal. 

Summarizing, it is possible to draw a well-defined boundary that separates the region of the parameter space where the system displays strong NMB. The simultaneous satisfaction of conditions (a) and (b) leads to ultra-NMB, but condition (b) alone is able to grant a relevant degree of nonmonotonicity even when the sparsity of the connections is violated.

% \tp{[[[EXPLAIN WHAT HAPPENS FOR LARGE k AND SMALL $\alpha$ AS FINITE SIZE EFFECT ONCE WE HAVE THE NEW HEATMAP N=50]]]}.

%It can be observed that, as a general trend, the larger is the fraction of asymmetric interactions $p_A$, the greater the maximum value of $\bar{k}$ for which a NMB is displayed. However, due to the very large synchronization time of the system in some regions of the parameter space, in our simulations, we have to keep the number of oscillators small. Consequently, because of the finite boundary conditions, when the angle of the CoV is small, it is not possible to increase the average number of neighbors beyond few units. This fact prevents us from confirming weather it still possible to recover the monotonic behavior by increasing the connectivity when $p_A$ approaches $1$. 

\section{A microscopic topological explanation for the emergence of NMB}\label{sec_conditions}

%We have already clarified that the connectivity pattern has to be at the same time both (a) sparse and (b) nonreciprocal to observe strong NMB. Nonetheless, a certain degree of nonmonotonicity persists when there is a large fraction of asymmetric interactions ($p_A$) regardless their density ($\bar{k}$).}
 
In this section we analyze in detail what happens at the local clusters when the parameters $\bar{k}$ and $p_A$ are varied. 
Our goal is to explain why and how some combinations of features (a) and (b) make the interplay between the mobility time scale and the time scale of local synchronization critically ineffective. To this end, the model under study is a very useful tool, different from the nearest-neighbors setting where $p_A$ depends on the number $k$ of neighbors with whom the units are allowed to interact. Indeed, through the parameters $R$ and $\alpha$, it is possible to tune both the density and the asymmetry of the interactions, respectively, while keeping the other constant.

Consider a static local cluster of nonsynchronized oscillators, firing at their out-neighbors (units having them inside their COV) and receiving from their in-neighbors (units inside their COV). 
We may ask ourselves what is needed for such a configuration to synchronize in finite time. Due to the asymmetry of the interactions, even if  each unit is receiving or sending signals to at least one other oscillator, it might be the case that some pathological configurations do not allow signal to flow throughout the entire cluster thus preventing it from synchronizing. 
We call such interaction patterns “frustrated configurations”: a setup that is not able to synchronize because of structure-related reasons. 

Consider, for instance, the case of two pairs of reciprocal neighbors firing at each other plus another oscillator that is not firing at anyone but is receiving from one element of each pair. These five units form a local cluster, but the signals interchanged between one pair do not affect the other in any way. Thus, they cannot synchronize. Every oscillator is receiving or sending signals to some units in the group, but still it is not possible for them to communicate at the cluster level.

\begin{figure}[htbp]
\begin{center}
\includegraphics[width=0.45\textwidth]{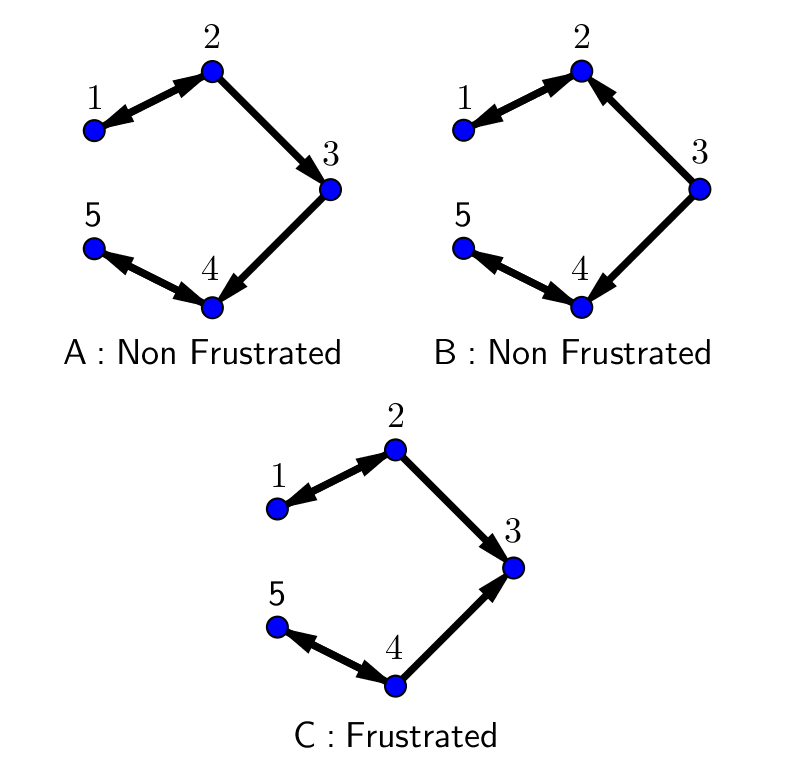}
\caption{Illustration of frustrated conditions. Cluster A is non frustrated because there is a path between all pairs of oscillators. Cluster B is non frustrated because there is an oscillator (3) from which there is a path to each oscillator of every pair without a path between them (pairs 1-4,1-5,2-4 and 2-5). Cluster C is frustrated because there is not an oscillator from which there is a path to each oscillator of a pair without a path between them (e.g., pair 2-4).}
\label{fig4}
\end{center}
\end{figure}

In topological terms, a configuration is frustrated if \emph{there does not exist an oscillator from which there is a (directed) path to every pair of oscillators which do not have a path between them}. Examples of this situation are depicted in Fig.~\ref{fig4}.

% fig 4: frustrated

%The abundance of frustrated configurations seems thus a determinant factor on the time the system takes to synchronize. 
In order to understand the impact of these configurations, imagine a cluster of oscillators laid out into a nonfrustrated configuration that is about to reach a synchronized state. If some change takes place in the internal interactions as a consequence of the motion (e.g. an oscillator exits any COV of the cluster) frustration may occur. Then, the phases that were converging towards a common value will move apart and the time they have spent together can be regarded as “wasted” in terms of the synchronization process. 
The worst scenario corresponds to the cluster having almost reached a coherent state when it changes into a frustrated configuration. Because of the peculiar characteristics of the pulse-coupling (with small enough refractory period, otherwise NMB does not appear \cite{perez2017control}), the achieved partial coherence will be wasted. Conversely, an optimal case would be the one in which a cluster becomes frustrated due to a change in the internal interactions just after synchronizing.
Indeed, if the cluster is already synchronized, such a change will not affect the achieved coherency. 
Any extra time spent by the oscillators in the same configuration once local synchronization has been attained does not help to enhance the coherence of the system. For lower velocities this happens more often and for longer periods, making the average (global) synchronization time longer.

% fig 5: heatmap3
\begin{figure}[htbp]
\begin{center}
\includegraphics[width=0.5\textwidth]{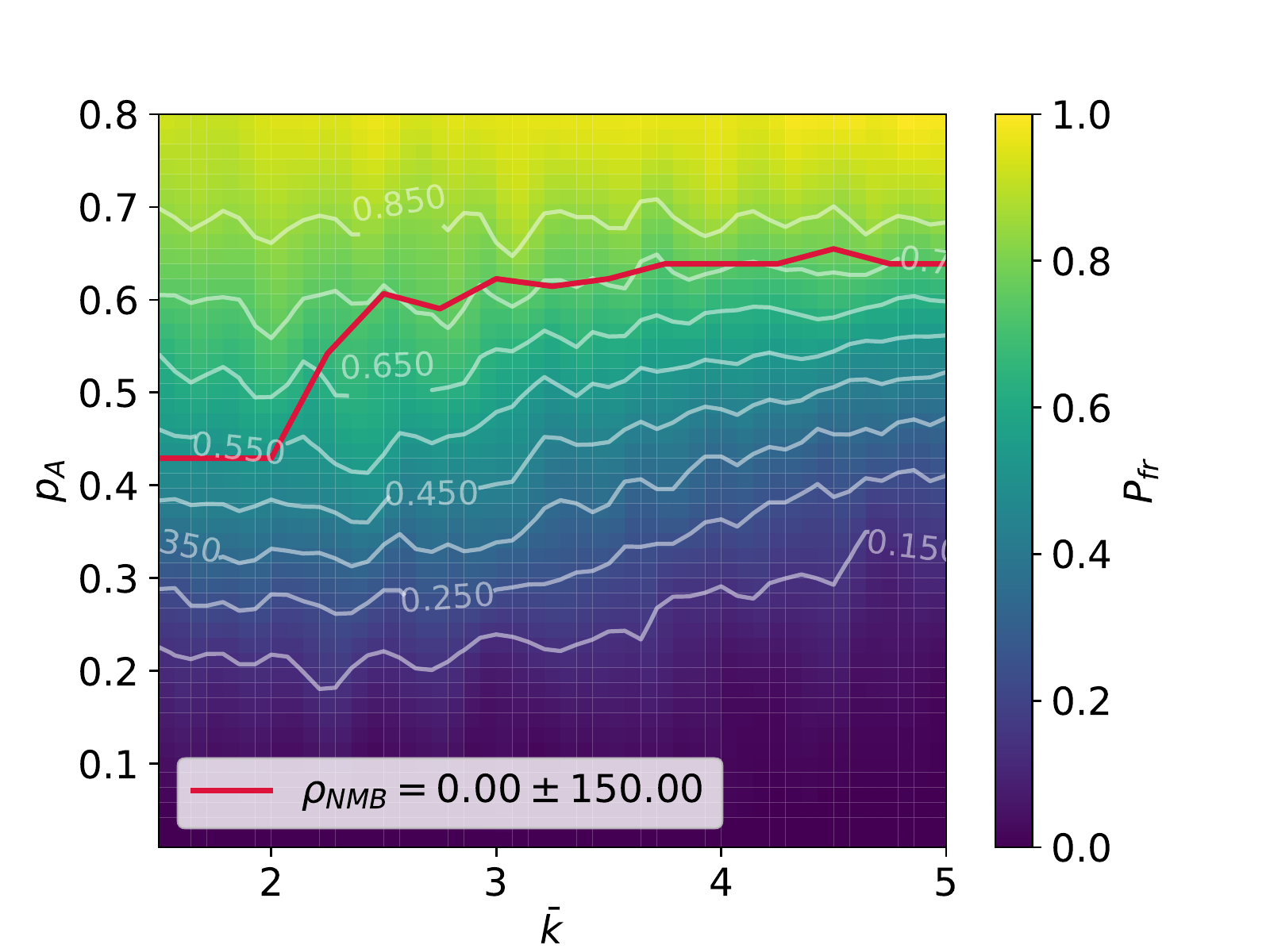}
\caption{$P_{fr}$ vs $p_{A},\bar{k}$. The probability of obtaining at least one frustrated cluster decreases by decreasing the asymmetry of the interaction pattern and, to a lesser degree, by increasing the connectivity. %The inset shows a point by point density comparison of the main figure ($P_{fr}(p_{A},\bar{k})$) relative to Figure \ref{fig3} ($\rho_{\text{Mm}}(p_{A},\bar{k})$), where the clearly defined functional relation among both NMB and $P_{fr}$ is apparent. 
The same boundary shown in Fig. \ref{fig3}(b) has been added as a guide for comparison. $P_{fr}$ averaged over 150 static interaction networks composed of $N=20$ oscillators. Heatmap interpolated using multiquadratic RBF.}% over the grey points in the background.}

\label{fig5}
\end{center}
\end{figure}

% fig 6: heatmap4
\begin{figure}[htbp]
\begin{center}
\includegraphics[width=0.5\textwidth]{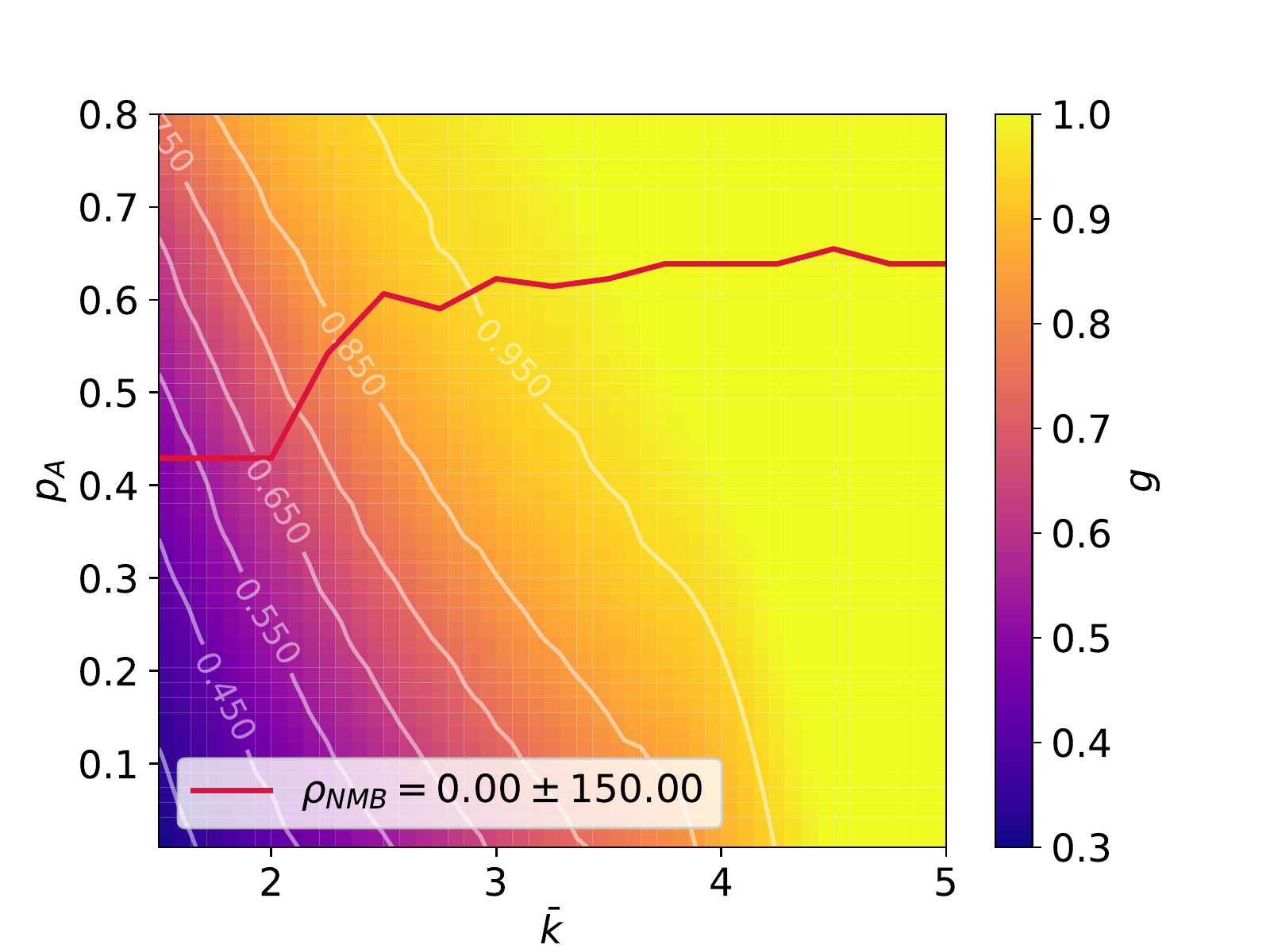}
\caption{Fraction of oscillators in the largest connected component $g$ against the expected fraction of asymmetric interactions, $p_{A}$, and the average number of neighbors, $\bar{k}$. Heatmap interpolated using multiquadratic RBF.} 
\label{fig6}
\end{center}
\end{figure}

The key factor driving synchronization at a global scale are hence non frustrated clusters that have reached local synchronization. For slow velocities, even if their overall fraction is small, for every “cluster change” the oscillators belonging to non frustrated (and already synchronized) clusters act as effective “spreaders” of synchronization across the system. However, as velocity is increased beyond  the point where nonfrustrated clusters cannot synchronize before a topological change occurs ($\nu>1$), the fraction of effective “spreaders” starts to decrease and synchronization is no longer promoted by mobility, leading to NMB. 

Synchronization time thus depends on the trade-off between two factors:
the presence of frustrated clusters in the system and the chances that nonfrustrated clusters have reached local synchronization before a change in the interaction pattern takes place. 

The first factor is a merely topological characteristic of the interaction pattern, which only depends on the interaction rules and not on the velocity of the oscillators. On the contrary, the second one is affected by mobility in a negative way: the higher the velocity, the smaller the chances that local clusters can synchronize. 

To quantify the first property, in Fig.~\ref{fig5} we show the probability of observing at least one frustrated configuration $P_{fr}$ in a randomly generated static connectivity pattern as a function of the fraction of asymmetric interactions, $p_A$, and the average number of neighbors $\bar{k}$. 

It can be deduced from the isolines that what affects $P_{fr}$ the most is $p_A$, while the only effect of $\bar{k}$ is that of slightly reducing the value $P_{fr}$ if increased at fixed $p_A$.

In this regard, the heatmap of $P_{fr}$ shows an overall resemblance to Fig.~\ref{fig3}(a), making this magnitude a good candidate to explain the transition between the monotonic and the nonmonotonic regions.

By overprinting the NMB boundary shown in Fig.~\ref{fig3}(b) on Fig.~\ref{fig5}, the relation between the boundary $\rho_{NMB}\approx 0$ (in red) and $P_{fr}$ is also clear. The boundary of the NMB occurs between isocurve $P_{fr}=0.55$ and isocurve $P_{fr}=0.75$. More precisely, it follows $P_{fr}=0.55$ when $\bar{k}\leq 2$, then moves to $P_{fr}=0.75$ as the connectivity increases.

This observation can be interpreted in terms of the general topology of the interaction pattern: Sparsity penalizes the achievement of synchronization in the intermediate velocity regime because global coherency through local synchronization is reached faster when there are large nonfrustrated clusters already synchronized (only one common phase for all the oscillators) than when there are several small clusters in the same situation. For that reason, the higher the connectivity, the higher $P_{fr}$ in order to make the local synchronization mechanism ineffective for intermediate velocities (which leads to NMB). Therefore, when the system is made up of separated small groups of connected oscillators, having at least one frustrated cluster in around $50-55\%$ of the static configurations is enough to trigger NMB. On the contrary, when the system is almost connected (the largest connected components includes more than $80\%$ of the oscillators) and the effect of mobility can be understood as that of a rewiring mechanism, a larger $P_{fr}$ is necessary in order to observe strong nonmonotonicity\footnote{Due to the limited number of oscillators, we cannot talk about a proper percolation transition.}. In Fig.~\ref{fig6}  we plotted the heatmap of the average fraction of oscillators belonging to the largest (weakly) connected component of the system to further clarify this point.

As a general conclusion, we can state that a certain amount of frustrated configurations are required for the system to display NMB. The value ranges between $P_{fr}=0.55$ and $P_{fr}=0.75$,  depending on the connectivity. It is not possible, at the present stage, to extend these results to other values of the rest of the parameters, and especially to other number of units in the system. However, most of the significant trends in the behavior of these systems do not change by changing $\epsilon$ or $N$ (see Appendix.~\ref{app:scaling_eps}). Even though we cannot make a strong claim stating that the precise amount of $P_{fr}$ required would not change under other conditions, it is very likely that the relation between frustration and nonmonotonicity would stay the same. In particular, it is important to stress, that lying below the percolation threshold is not a necessary condition for the appearance of NMB: If the fraction of asymmetric interaction is large enough, the existence of frustrated configuration enhances nonmonotonicity even when all the oscillators belong to the same cluster.

%Beyond this boundary, increasing the mobility of the oscillators has the effect of narrowing the fraction of “spreaders” to the point that there is no coalescence of small groups of synchronized units into larger ones. The system is stuck in a situation of achieving little local coherence just to lose it immediately after.

%\tp{COMMENT ON SMALL ALPHA, IN THIS CASE, THERE ARE STILL GROUPS OF OSCILLATORS, BUT THE STATISTIC IS CLEARLY LOST DUE TO THE PERCOLATION PROBLEM}
%a point where it is insufficient to force the entire system to synchronize.

%For even higher velocities, the notion of local cluster is lost, as all interactions are effectively independent. Then, all units contribute to the synchronization process every time their signal reaches another unit and whenever they adjust their phases upon a firing event from oscillators inside their COV. Thus the system reaches global synchronization in a increasingly faster way.

\section{Conclusions}\label{sec_conclusions}
%FINISH!
The present work constitutes another step forward to understand the peculiar phenomena of the prevention of synchronization of a group of mobile PCOs by tuning their velocities, presented in \cite{PSD13} and further studied in other works \cite{PRG15,7277178,perez2017control}. 

In this research, we have first shown how two conditions are needed for this phenomenology to appear, mainly a sparse connectivity pattern and asymmetric interactions. We have confirmed a control parameter $\nu$, a quantity expressed in terms of the quotient among the time a typical cluster of oscillators takes to synchronize and the average time spent for these clusters to suffer a connectivity change due to mobility. We furthermore have put forward a microscopic explanation to show how the appearance of frustrated clusters is the most likely explanation behind the nonmonotonic dependency of the average synchronization time on the velocity in systems of moving PCOs.

Frustration can be regarded as an emerging property of the connectivity pattern that solely depends on the interaction rules implemented in each model, not on their dynamics. This leads us to believe that any model of the same class (moving PCOs) might display exactly the same behavior.

Additionally, we have proposed two metrics that allow one to (1) determine if the system can display NMB and (2) predict for which values of the velocity such behavior may occur. 
The first metric is the probability of finding a frustrated configuration $P_{fr}$. Although we are not able, at the present moment, to suggest a precise value of $P_{fr}$ above which NMB will be observed, we can confidently state that if $P_{fr}$ is high enough, NMB will surely appear. The threshold value is affected by the concurrence of other factors, such as the size of frustrated clusters or the existence of synchronizable subclusters within one frustrated group. The relevance of these secondary factors is very difficult to analyze because of finite size effects. Further efforts need to be devoted to determining the precise necessary conditions for NMB in terms of the values of the parameters of the model. 

The second metric, $\nu$, had already been conceptually introduced in \cite{PSD13}. In this paper, we confirmed that it is a general fact that, above $\nu=1$, when the typical time for local synchronization is larger than the average time between two topological changes, the behavior of the system may deviate from monotonicity.

It must be noted that the phenomenology studied here may have large relevance due to their application in swarms of autonomous robotic vehicles, as its appearance has been reproduced in experimental settings \cite{perez2017control}. The present work sheds light on explaining the appearance of this intriguing emerging behavior and, moreover, helps in identifying possible general features that might be not only restricted to mobile pulse-coupled oscillators but can be applied more generally to wider sets of models subject to discrete, nonlinear, firing processes. The explanation provided in this work constitutes thus solid ground from which to test this hypothesis on other models, equally nonlinear, proposed in the literature, which might be especially relevant for new technological applications in robotics.

\acknowledgments The authors acknowledge support from
Ministerio de Econom\'{\i}a y Competitividad of Spain Projects
No. FIS2012-38266-C02-02 and No. FIS2015-71582-C2-2-P
(MINECO/FEDER); and Generalitat de Catalunya Grant No.
2014SGR608.
\appendix
%%%%%%% Apendices %%%%%%%%%%%

\section{Average Neighbours (Finite Boundary Conditions Effects)}\label{app:scaling_k}

This Appendix shows the estimation of the estimation of the average number of neighbors that we used to take into account for the finite boundary conditions.
With periodic boundary conditions, the average fraction of neighbors is equal to the fraction of area covered by the cone of vision,
\begin{align}
\frac{\bar{k_{0}}}{(N-1)}=\frac{\alpha R^{2}}{2L^{2}}.
\end{align}

\begin{figure}[htbp]
\begin{center}
\includegraphics[width=0.5\textwidth]{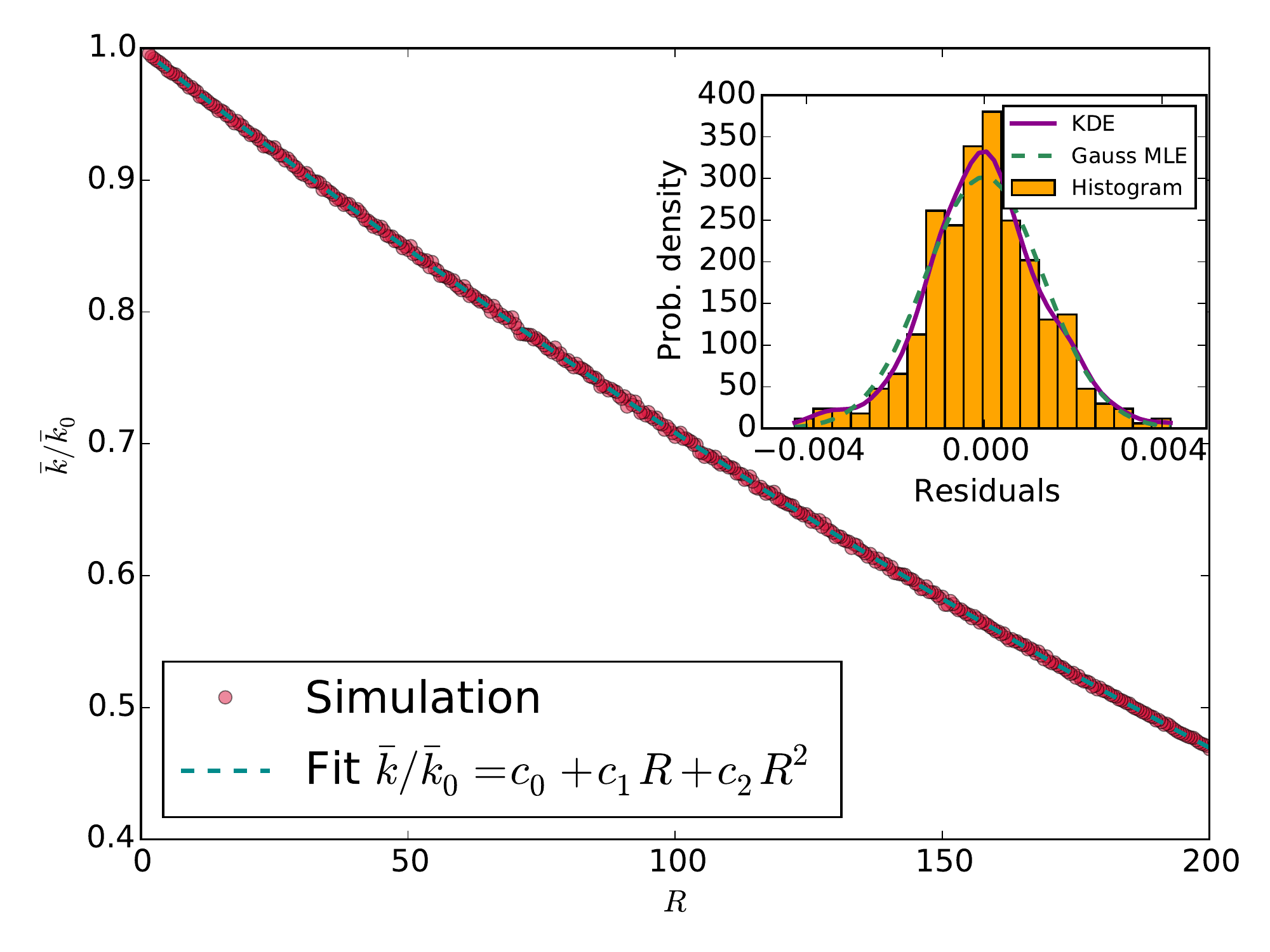}
\caption{Average number of neighbors, $\bar{k}$, with finite boundary conditions relative to the theoretical value of the average number of neighbors with periodic boundary conditions, $\bar{k}_0$, as a function of the COV radius $R$, including results of the simulation and a quadratic fit. The inset shows the residuals of the fit together with the maximum likelihood fit to a normal distribution and kernel density estimator (KDE). $\bar{k}$ is computed by averaging over 10000 static realizations.}
\label{fig_app_finite_conds}
\end{center}
\end{figure}

Taking into account finite boundary conditions, the average number of neighbors, $\bar{k}$, is smaller than $\bar{k_{0}}$ because the area of the cone can be partially out of the space. The orientations are uniformly distributed and hence $\frac{\bar{k}}{\bar{k_{0}}}$ does not depend on $\alpha$, but depends on $R$. Fig.~\ref{fig_app_finite_conds} shows the corresponding fit leading to the following expression for the average number of neighbors with finite boundary conditions:
\begin{align}
\begin{split}
\frac{\bar{k}}{\bar{k_{0}}} &= \frac{\bar{k} }{\frac{\alpha R^{2} (N-1)}{2L^{2}}} = c_0 + c_1 R + c_2 R^2, \\
c_0 = 0.9996, &\qquad c_1 =-3.2 \cdot 10^{-3}, \qquad c_2 = 2.6 \cdot 10^{-6}.
\end{split}
\end{align}

The fit has been performed using Ridge regression with basis expansion $(R,R^2)$ obtaining very accurate results, as shown from the Gaussianity of the residuals (See inset in Fig.~\ref{fig_app_finite_conds} and the good result of the reduced $r^2$ value: $r^2>1-1\cdot 10^{-5}$.)

\section{Analytical calculation of the Neighbour Time for $\alpha=\pi$}\label{app:pi_time}

In this Appendix we perform the analytic calculation of the average time between two sequential changes in the interaction pattern, $T_{C}$, for $\alpha=\pi$, although the procedure can be generalized for an arbitrary $\alpha$. 

Let $t_{C}$ be the average time an oscillator needs in order to stop having another one inside its COV, and thus to stop having it as its neighbor. We have $T_{C}=t_{C}/N$, where $N$ is the number of oscillators in the system.

Consider the representation of Figure~\ref{fig_app_tneigh}: The oscillator with the cone $O_{1}$ is located in the origin oriented with velocity $v_{1}=V\hat{\mathbf{i}}$ and the oscillator $O_{2}$ is located in some point of the semicircle (at distance $R$ and angle $\Phi\in[-\frac{\pi}{2},\frac{\pi}{2}]$ from $O_{1}$). Given that both oscillators have the same modulus velocity $V$, the positions considered for $O_{2}$ are the only ones from which this oscillator can enter inside the cone. Let $\gamma$ be the random orientation of $O_{2}$ and $\mathbf{v}_{2}=V(\hat{\mathbf{i}}\cos\gamma+\hat{\mathbf{j}}\sin\gamma)$ its velocity. Hence the relative velocity $\mathbf{v}$ is
\begin{align}
\mathbf{v}=\mathbf{v}_{2}-\mathbf{v}_{1}=V(\hat{\mathbf{i}}(\cos\gamma-1)+\hat{\mathbf{j}}\sin\gamma).
\end{align}

Consider the horizontal relative distance $x(t)$ and the vertical relative distance $y(t)$ as a function of time $t$:

\begin{figure}[htbp]
\begin{center}
\includegraphics[width=0.5\textwidth]{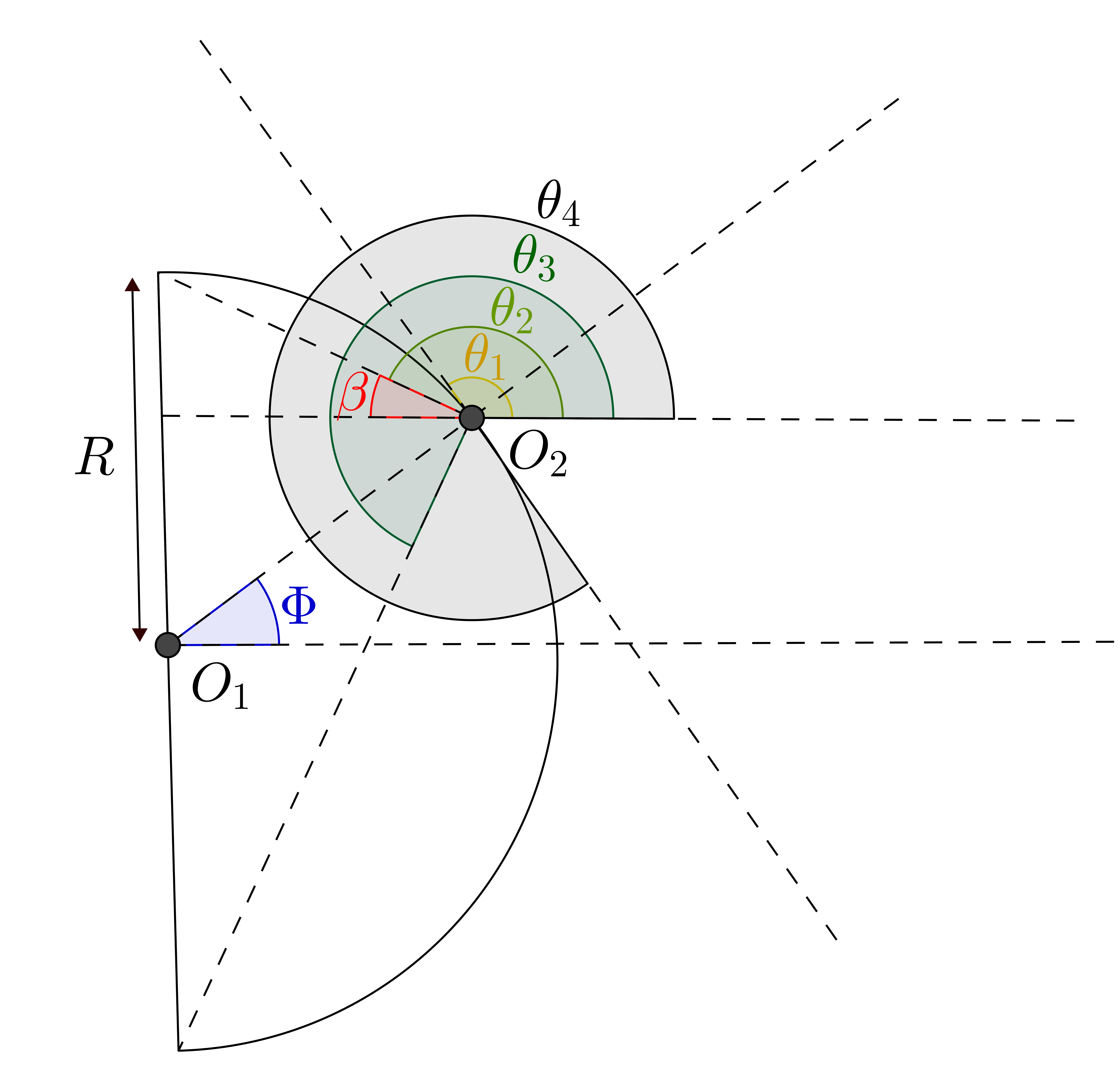}
\caption{Representation of the integral variables: $\Phi$ is the initial angular coordinate of oscillator $O_{2}$; angles $\theta_{1}$, $\theta_{2}$, $\theta_{3}$, and $\theta_{4}$ are the extremes of integration and $\beta$ is an instrumental variable in the calculation.}
\label{fig_app_tneigh}
\end{center}
\end{figure}

\begin{align}
\begin{split}
x(t)&=R\cos\Phi+tV(\cos\gamma-1)\\
y(t)&=R\sin\Phi+tV\sin\gamma.
\end{split}
\end{align}
Consider the variable
\begin{align}\label{theta_var}
\theta=\tan^{-1}\left (\frac{\sin\gamma}{\cos\gamma-1}\right)
\end{align}
which is the polar coordinate of the relative velocity. Notice that, with the notation introduced in Fig.~\ref{fig_app_tneigh}, $O_{2}$ enters inside the cone if $\theta\in(\theta_{1}(\Phi),\theta_{4}(\Phi))$. Therefore, the integral we must solve to find $t_{C}$ is
\begin{align}\label{eq_int_time}
t_{C}= \langle T(\theta,\Phi) \rangle = \frac{1}{\pi}\int^{\frac{\pi}{2}}_{-\frac{\pi}{2}}d\Phi\frac{1}{\pi}\int^{\theta_{4}(\Phi)}_{\theta_{1}(\Phi)}d\theta \,T(\theta,\Phi)
\end{align}
where $T(\theta,\Phi)$ is the time that $O_{2}$ needs to leave the cone.

Notice that if $\theta\in(\theta_{1}(\Phi),\theta_{2}(\Phi))$, then $T(\theta,\Phi)\equiv T_{1}$ satisfies
\begin{align}
(x(T_{1}))^{2}+(y(T_{1}))^{2}=R^{2}
\end{align}
and $T_{1}> 0$.

If $\theta\in(\theta_{2}(\Phi),\theta_{3}(\Phi))$, then $T(\theta,\Phi)\equiv T_{2}$ satisfies
\begin{align}
x(T_{2})=0.
\end{align}

And finally if $\theta\in(\theta_{3}(\Phi),\theta_{4}(\Phi))$, then $T(\theta,\Phi)\equiv T_{3}$ satisfies
\begin{align}
(x(T_{3}))^{2}+(y(T_{3}))^{2}=R^{2}
\end{align}
and $T_{3}> 0$.

Therefore, integral~\eqref{eq_int_time} becomes
\begin{align}\label{eq_int_time2}
\begin{split}
t_{C}&=\frac{1}{\pi}\int^{\frac{\pi}{2}}_{-\frac{\pi}{2}}d\Phi\,\frac{1}{\pi}\int^{\theta_{2}(\Phi)}_{\theta_{1}(\Phi)}d\theta\, T_{1}+\\
&\frac{1}{\pi}\int^{\frac{\pi}{2}}_{-\frac{\pi}{2}}d\Phi\frac{1}{\pi}\int^{\theta_{3}(\Phi)}_{\theta_{2}(\Phi)}d\theta \, T_{2}+\frac{1}{\pi}\int^{\frac{\pi}{2}}_{-\frac{\pi}{2}}d\Phi\frac{1}{\pi}\int^{\theta_{4}(\Phi)}_{\theta_{3}(\Phi)}d\theta \, T_{3}
\end{split}
\end{align}
Now we need to determine $\theta_{1}$, $\theta_{2}$, $\theta_{3}$, and $\theta_{4}$. Consider
\begin{align}
\beta=\tan^{-1}\left(\frac{1-\sin(|\Phi|)}{\cos\Phi}\right).
\end{align}

We have
\begin{align}
\begin{split}
\theta_{1}&=\Phi+\frac{\pi}{2}\\
\theta_{2}&=\pi-\beta\\
\theta_{3}&=\theta_{2}+\frac{\pi}{2}=\frac{3\pi}{2}-\beta\\
\theta_{4}&=\Phi+\frac{3\pi}{2}.
\end{split}
\end{align}
Moreover, from Eq.~\eqref{theta_var},
\begin{align}
\cos\gamma-1=\frac{\sin\gamma}{\tan\theta},
\end{align}
\begin{align}
\sin\gamma=\frac{-2\tan\theta}{\tan^{2}(\theta)+1}.
\end{align}

Now we use the previous expressions to isolate $T_{i}$, discarding the null solutions. We obtain

\begin{align}\label{Times}
\begin{split}
T_{1}&=\frac{R(\sin\Phi\tan\theta+\cos\Phi)}{V},\\
T_{2}&=\frac{R\cos\Phi(\tan^{2}(\theta)+1)}{2V},\\
T_{3}&=\frac{R(\sin\Phi\tan\theta+\cos\Phi)}{V}.
\end{split}
\end{align}

Finally, we substitute expressions \eqref{Times} into integral~\eqref{eq_int_time2}. The resulting integral is solvable analytically and we obtain the expression
\begin{align}
t_{C}=\frac{3R}{\pi V}
\end{align}
and hence
\begin{align}\label{final_t_neigh}
T_{C}=\frac{3R}{\pi V N}.
\end{align}

\section{Appearance of NMB above $\nu \geq 1$}\label{app_nu1}

In this Appendix we provide further proof that $\nu$ is the appropriate control parameter and that NMB appears when $T_C$ equates $T_L$, that is, when $\nu=1$. To check for this, we have proceeded to fit the curves in Fig.~\ref{fig12} according to the relation $\langle T_{\text{sync}} \rangle = A\nu^{\gamma}$ taking an increasing number of points in the interval $\nu \in (0,\nu_{\text{max}}]$. The fits have been performed using a standard least squares linear fit on the log transformed variables $\ln \langle T_{\text{sync}} \rangle \sim \gamma \ln \nu + \ln A$.

We proceed to plot two standard metrics relating to the best fit of each line, its average squared error $\delta^2$ and explained variance $\rho_{\text{var}}$ defined, respectively, as
\begin{align}
\begin{split}
\delta^2 &= \frac{1}{N_{\text{points}}} \sum_{i}^{N_{\text{points}}}  \left (\langle T_{\text{sync}} \rangle_i - A\nu^{\gamma}_i \right)^2.  \\
\rho_{\text{var}} &= 1 - \frac{\left (  \langle T_{\text{sync}} \rangle - A\nu^{\gamma} - \overline{\langle T_{\text{sync}} \rangle - A\nu^{\gamma} } \right)^2}{\left( \langle T_{\text{sync}} \rangle - A\nu^{\gamma}\right)^2} \\
 \overline{\langle T_{\text{sync}} \rangle - A\nu^{\gamma} } &\equiv \frac{1}{N_{\text{points}}} \sum_{i}^{N_{\text{points}}}  \langle T_{\text{sync}} \rangle_i - A\nu^{\gamma}_i.
\end{split}
\end{align}

\begin{figure}[htbp]
\begin{center}
\includegraphics[width=0.5\textwidth]{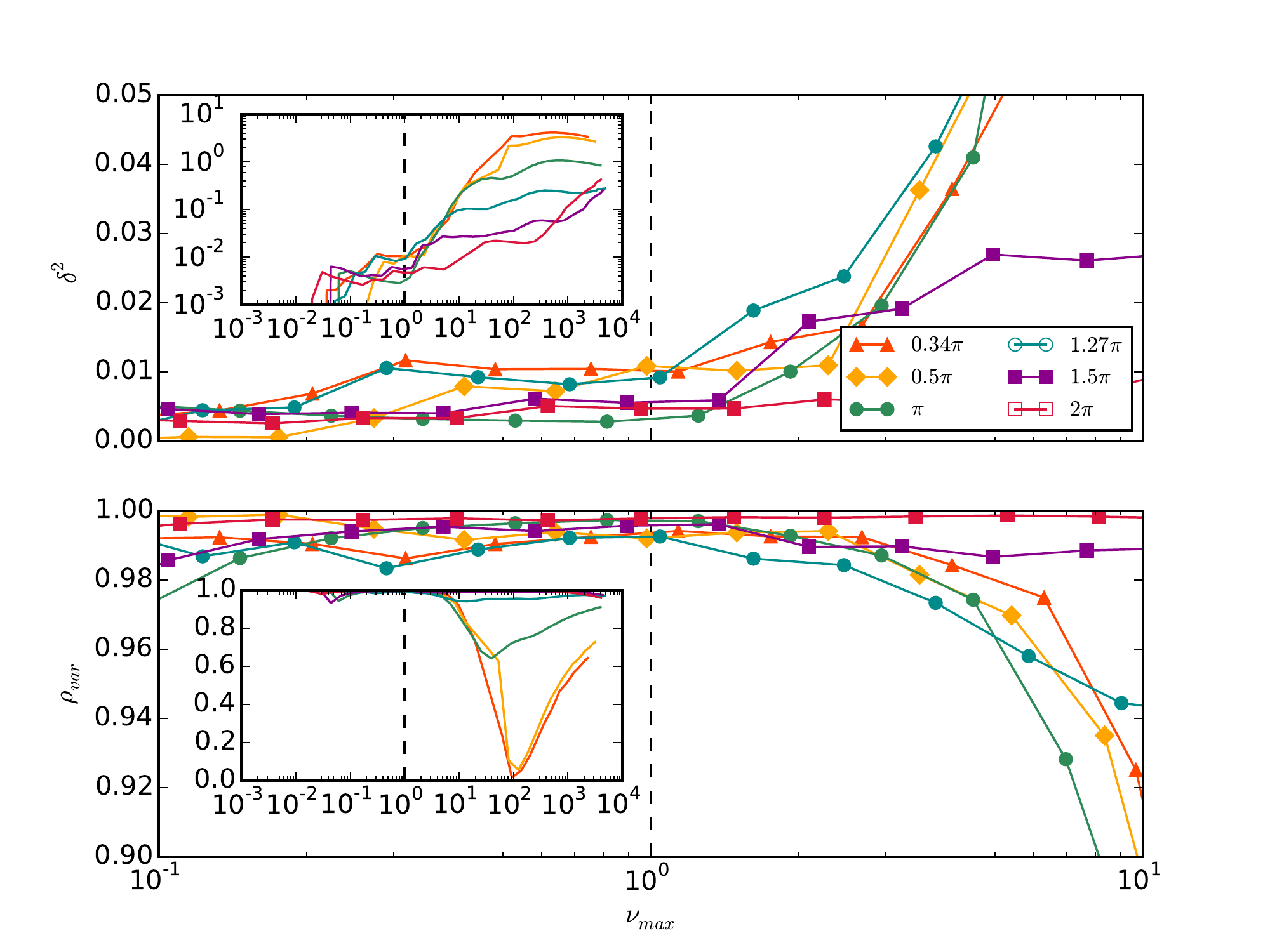}
\includegraphics[width=0.5\textwidth]{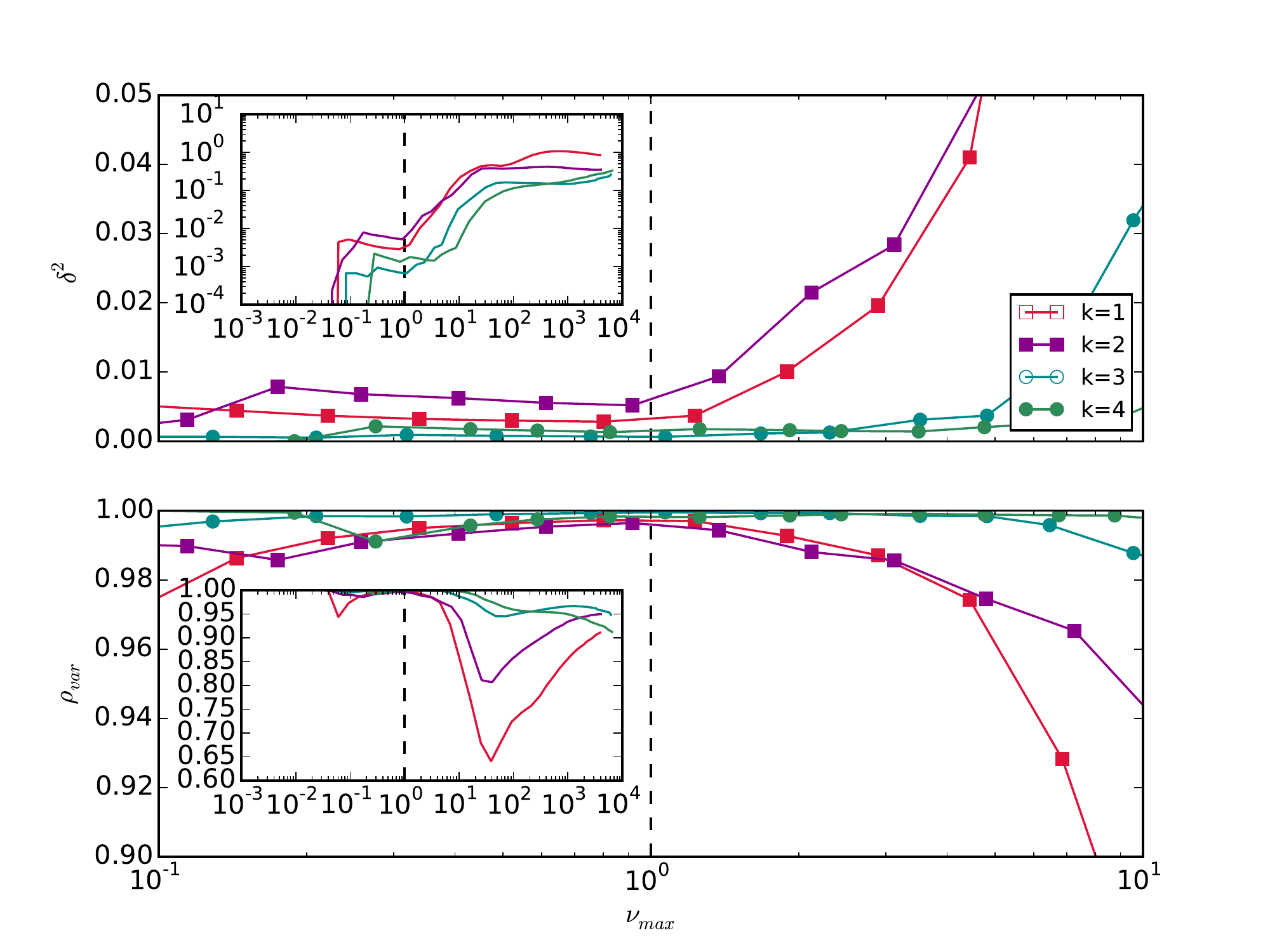}
\caption{$\delta^2$ and $\rho_{\text{var}}$ vs $\nu$ corresponding to curves in Fig.~\ref{fig12} around the area of interest $\nu_{\text{max}}=1$. The displaying of the NMB (if present) discussed appears in $\nu>1$ as seen by the decrease in the quality of the power law fit. Insets show log log plots over the complete set of $\nu_{\text{max}}$ values.}
\label{fig_app_nu1}
\end{center}
\end{figure}

As seen in Fig.~\ref{fig_app_nu1}, whenever NMB distinctively appears, the quality of the fit significantly decreases starting at the point $\nu_{\text{max}} \simeq1$. The decrease in fit quality with $\nu_{\text{max}}$ for the cases where no NMB is present is due to the appearance of the fast velocity regime, where $\langle T_{sync} \rangle$ becomes independent of $\nu$ as discussed earlier.

%Concerning the values of $\gamma$ emerging from the fit, although unimportant for 

\section{Average synchronization time for varying $\epsilon$ and $N$}\label{app:scaling_eps}

In this Appendix we show that neither the coupling parameter $\epsilon$ nor the number of units, $N$, affects the appearance of nonmonotonic trends in the emergence of synchronization through mobility characterized in Sec. \ref{sec_model}. As expected, if the interaction pattern satisfies conditions (a) and (b), the system deviates from the monotonic behavior starting from $\nu=1$, for any (small) value of $\varepsilon$ and any $N$.

\begin{figure}[htbp]
\begin{center}
\includegraphics[width=0.5\textwidth]{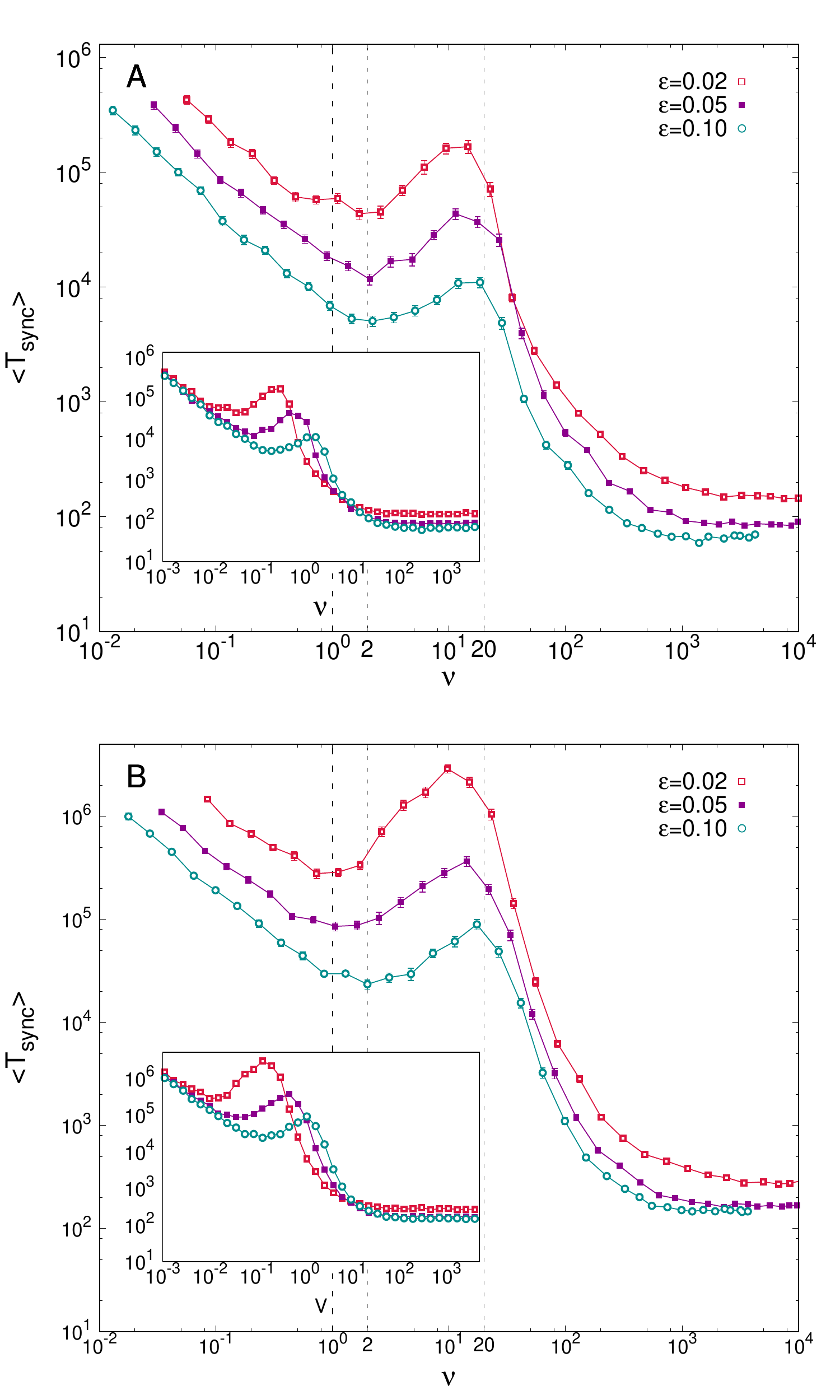}
\caption{$\langle T_{\text{sync}}\rangle$ vs $\nu$. We fix (a) $\bar{k}=2$ and (b) $\bar{k}=1$. The display of the NMB discussed in Sec. \ref{sec_model} does not depend on the coupling parameter $\epsilon$. $\langle T_{\text{sync}}\rangle$ was calculated by averaging over 75 realizations with $N=20$ and $\alpha=\pi$.}
\label{fig_app_eps_scaling}
\end{center}
\end{figure}

\begin{figure}[htbp]
\begin{center}
\includegraphics[width=0.5\textwidth]{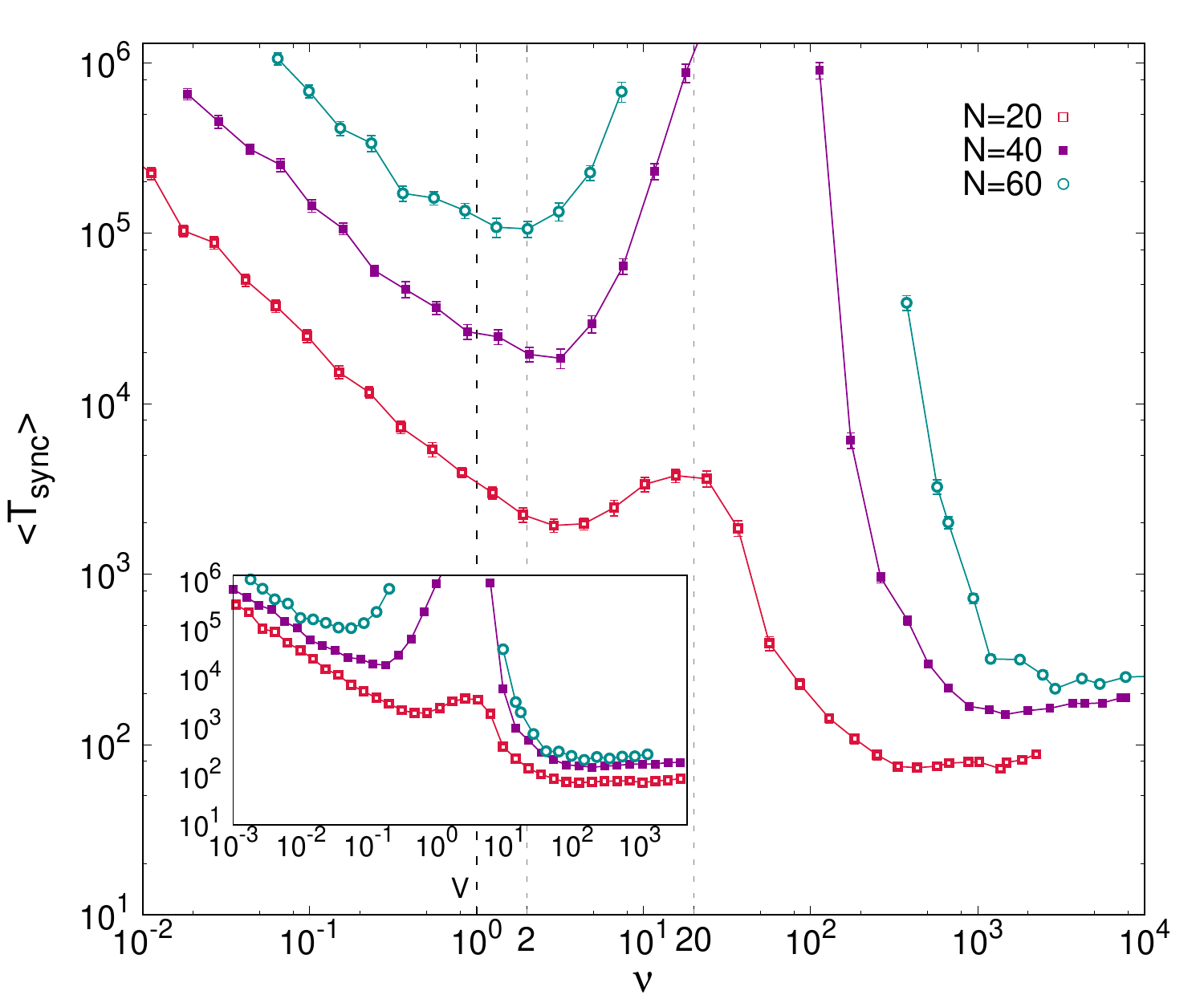}
\caption{$\langle T_{\text{sync}}\rangle$ vs $\nu$. The displaying of the NMB discussed in Sec. \ref{sec_model} does not depend on the number of oscillators $N$. $\langle T_{\text{sync}}\rangle$ was calculated averaging over 75 realizations with $\bar{k}=2$, $\varepsilon=0.2$ and $\alpha=\pi$.}
\label{fig_app_N_scaling}
\end{center}
\end{figure}

In Figs.~\ref{fig_app_eps_scaling} and~\ref{fig_app_N_scaling} we plot the average synchronization time as a function of the control parameter $\nu$ with varying $\varepsilon$ and $N$ respectively, for a fixed value of the geometric parameters of the COV, $\alpha=\pi$ and $\bar{k}=1,2$. Again, we observe an alignment of all the curves under the $\nu$-rescaling. 

Therefore, we can conclude that $\nu$, the ratio between the two time scales of local synchronization and topological change, is the appropriate control parameter for this class of systems.

\bibliography{library}
\end{document}